# Electronic and optical properties of SrTiO$_3$ perovskite using semi-local and hybrid first principles density functional theory

M. Kar [a]

a. Department of Engineering Materials, School of Engineering, University of Southampton, United Kingdom

M.Kar@soton.ac.uk

**Abstract**

We study the crystal structures, electronic and optical properties of SrTiO$_3$ perovskite in 3 phases; namely cubic, orthorhombic, and tetragonal using first principles Density Functional Theory (DFT) calculations. We used the Generalised Gradient Approximation (GGA) within the DFT formalism to optimise the crystal structures of SrTiO$_3$ in the 3 phases and employed the hybrid functional Heyd Scuserof Ernzerhof (HSE) to predict the energy band gap of these structures. From our calculations, we predicted the ground state lattice constant of 3.96 angstroms for the cubic phase; close to the experimental lattice constant of 3.92 angstrom and energy band gap of 3.3 eV for cubic SrTiO$_3$ close to it's experimental band gap of 3.25 eV. We further employed the HSE formalism to predict the band gaps in the orthorhombic and tetragonal phases; however there has been no experimental literature to compare the same with.

**Introduction**

Complex correlated oxides in general have several interesting properties such as high Tc superconductivity, multiferroicity, colossal magnetoresistance and ferroelectricity. Most interesting complex oxides have a perovskite structure [1-4]; hence SrTiO$_3$ is considered to be an ideal starting point for their study. Strontium titanate (SrTiO$_3$) is an important perovskite oxide that has been studied extensively over the past years due to it's technological importance. [5-7] Several modern concepts of condensed matter and physics of phase transition have been studied while investigating this unique material. Among it's wide range of applications, includes optical switches, grain boundary barrier layer capacitors, oxygen gas sensors, wide band gap semiconductivity, etc. [8-16]

Over the past years, the structural, electronic, optical, and elastic properties of SrTiO$_3$ have been investigated thoroughly both experimentally [17-23] and theoretically [24-27]. The theoretical description of this material is however still an active area of research. Conventionally, theoretical computation of solid materials face difficulty in predicting their correct band gap and the associated electronic properties from first principles. Density functional theory (DFT) additionally with the Coulomb interaction formalism (DFT+U) [28-31] has been successful in obtaining the correct energy band gaps, however only for correlated and localised electrons in transition and rare earth metal oxides.

The underestimation of the energy band gaps and the associated electronic properties is typically owed to the inadequacies of the current density functional potentials for the description of ground state properties of semiconductors. In some cases, methods beyond DFT also have proven to be unsuccessful towards predicting the electronic properties of semiconducting materials [32-34]. In general, the perovskite materials exist in 3 phases dependent on the temperature. At low temperatures <100K, these exist in the orthorhombic phase, at temperatures between 100K and 300K, these exist in the tetragonal phase, and at high temperature above 300K, cubic phase of the perovskites are found to be the most stable. [35-37] The major motivation of this work is to study the crystallography predict the energy band gap of SrTiO$_3$ in all these 3 phases using first principles DFT calculations. We are able to predict band gaps close to the experimental band gap by using the HSE functional that mixes certain amount of Hartree-Fock exchange to the standard DFT functional that is known to improve the band gaps of solids.

**Computational Details**

All calculations are done using plane-wave Density Functional Theory (DFT) simulations as implemented within the Vienna Ab-initio Software Package (VASP). The initial geometries of the perovskites in all the 3 phases (cubic, orthorhombic and tetragonal) are taken from Materials Project website and are given in Table 1. The geometries of the perovskites are optimised using the Perdew-Burke-Ernzerhof (PBE) functional using a cut off energy of 520 eV. Plane wave basis sets with ultrasoft pseudopotentials as implemented within the VASP package have been employed for Sr, Ti, and O. The sampling of the Brillouin zone has been done using a k-point mesh of 5x5x5. The energy band gaps are calculated using both the PBE and the hybrid HSE functionals. The parameters have been tested computationally for convergence.

**Results and Discussions**

The geometries of the SrTiO3 perovskites are optimised using the Broyden-Fletcher-Goldfarb-Shanno (BFGS) algorithm as implemented within the VASP package. The optimised crystals of SrTiO3 perovskite is seen to be undistorted in all the 3 phases. For the cubic and the orthorhombic crystal phases, however, the lattice parameters are seen to decrease after optimisation, thus reducing the cell volume. For the tetragonal crystal structure, the lattice parameters increase by little amount post optimisation, thus increasing the cell volume by a small percentage. The initial and the final lattice parameters, lattice volume and the percentage differences are shown in table 1.

|  | a(ini) Å | b(ini) Å | c(ini) Å | a(opt) Å | b(opt) Å | c(opt) Å | Volume(ini) Å$^3$ | Volume(final) Å$^3$ | %age diff. |
|---|---|---|---|---|---|---|---|---|---|
| cubic | 4.2556 | 4.2556 | 4.2556 | 3.9627 | 3.9627 | 3.9627 | 77.07 | 62.23 | 19.2 |
| orthorhombic | 5.8471 | 5.9123 | 8.2983 | 5.605 | 5.592 | 7.918 | 286.87 | 248.19 | 13.48 |
| tetragonal | 5.56773 | 5.56773 | 7.90711 | 5.589 | 5.589 | 7.937 | 245.11 | 247.98 | 1.17 |

**Table 1: The initial and final lattice parameters obtained from DFT calculations in all 3 phases.**

The energy band gaps of the perovskite have been predicted by using both the PBE as well as the HSE functional in all the 3 crystal phases. PBE functional tends to underestimate the energy band gap of periodic solids. The band gap calculation of solids using the PBE functional comes with a phenomenon that is called self-interaction error, which arises in the occupied states in DFT and in the unoccupied states in Hartree-Fock calculations. These spurious self interaction error effects in the occupied states in semi local DFT delocalises these occupied states even further, thus shooting up the energy and reducing the band gap. Hartree-Fock theory has the same problem, however, in the unoccupied states; these states are over delocalised and hence pushed up, thus increasing the band gap. Empirically, it's found that mixing a semi-local DFT functional with a small amount of exact Hartree-Fock exchange equalises the self interaction error of the top occupied band and the lowest unoccupied band, thus making the band gap close to experimental band gap. The PBE and the HSE band gaps of the perovskite in all 3 phases are shown in table 2.

| SrTiO$_3$ | PBE predicted band gap (eV) | HSE predicted band gap (eV) | Band gap predicted from literature | Experimental band gap (eV) |
|---|---|---|---|---|
| Cubic | 1.88 | 3.3 | 1.9 | 3.25 |
| Orthorhombic | 1.90 | 3.12 | - | - |
| Tetragonal | 1.857 | 3.4 | - | - |

**Table 2: The DFT predicted band gaps using both PBE and HSE band gaps and their comparison with the experimental band gap. '-' denotes no data is available from literature.**

The 0K phase stability of the perovskite can be determined approximately from the DFT total energies as shown in figure 2 . Conventionally, the room temperature phase stability of the perovskites can be estimated using the empirical Goldschmidt tolerance factor t. Goldschmidt tolerance factor is an empirical descriptor-based method used for predicting the stability of the perovskites. This method is based on the chemical composition of the perovskite ABX$_3$ and the ionic radii of the A, B, and t X ions. t is calculated using the formula:

$$t = \frac{r_A + r_X}{\sqrt{2}(r_B + r_X)}$$

Here, $r_A$, $r_B$, and $r_X$ are the radii of the A, B, and X ions. t=1 indicates a perfect, cubic perovskite. A shift of the value from t=1 indicates a distortion in the structure. A value less than 1 implies an orthorhombic structure, whereas t > 1 implies a tetragonal/hexagonal structure. The t value of SrTiO3 is calculated to be 0.97, thus implying a distorted cubic room temperature stable crystal phase. The ionic radii used for the calculation of the Goldschmidt factor has been given in table 3.

| Ion | Ionic radius (Å) |
|---|---|
| $Sr^{2+}$ | 1.13 |
| $Ti^{4+}$ | 0.68 |
| $O^{2-}$ | 1.4 |

**Table 3: The ionic radii of the elements used for the calculation of the Goldschmid tolerance factor.**

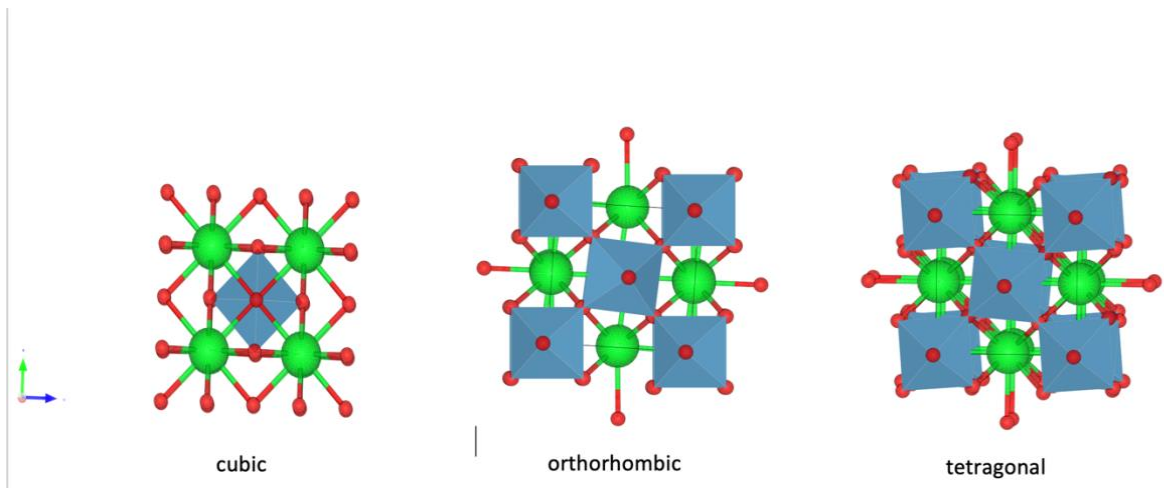

**Figure 1: The crystal structure of SrTiO₃ in 3 phases.**

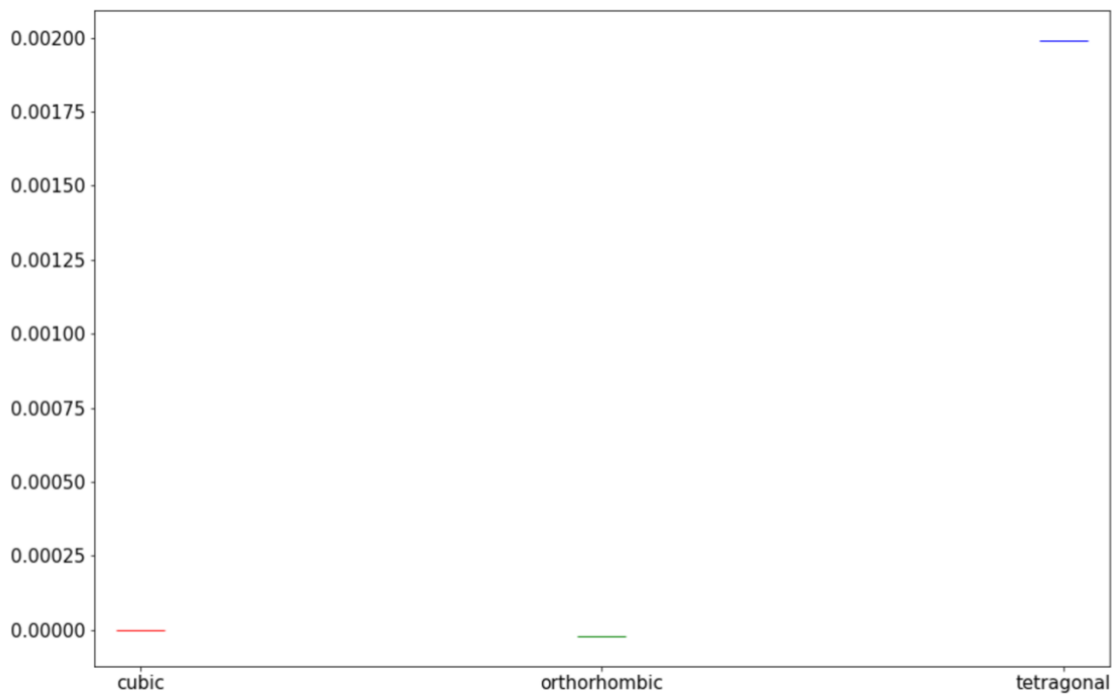

**Figure 2: The 0K phase stability of SrTiO₃ using Density Functional Theory. The cubic and orthorhombic phases are seen to be more stable than the tetragonal phase.**

## Conclusions

The crystal structure, electronic properties and crystallographic phase stabilities of SrTiO3 perovskite has been studied using first principles DFT calculations. From our findings, we conclude that this perovskite material can be used as a potential wide band gap semiconductor in all 3 possible crystallographic phases. Experimental data is available so far for the cubic crystal phase only. However, from our study, the other two phases are also predicted to be quite promising for applications as wide band gap semiconductors.

## Acknowledgements

Both the named authors have equal contribution towards the planning, discussion and calculations involved within the project. The authors thank the supercomputing cluster IRIDIS5 within the University of Southampton for providing the necessary computational resources. The authors also thank Dr P. Ghosh and his group at Indian Institute of Science Education and Research (IISER) Pune for rendering immense support and discussion towards the completion of this work.

## References

1. H. Oata et al., Nature, 6, 129, (2007).
2. W. Wunderlich et al., Physica B, 404, 2202, (2009).
3. A. D. Caviglia et al., Nature, 456, 624, (2008).
4. Y. J. Chang et al., Phys. Rev. B, 81, 235109, (2010).
5. E. Heifets et al., J. Phys. Condens. Mater., 18, 4845, (2006).
6. Y. X. Wang et al., Solid State Comm., 120, 133, (2001).
7. M. E. Lines and A. M. Glass, Principles and Applications of Ferroelectrics and related Materials, (Clarendon Press, Oxford), 1997.
8. J. D. Wilk et al., J. Appl. Phys., 89, 5243, (2001).
9. R. I. Eglitis et al., Ceramics International, 30, 1989, (2004).
10. S. Piskunov et al., Comp. Mat. Sci., 29, 165, (2004).
11. S. Tinte et al., Phys. Rev. B, 59, 1959, (1998).
12. Y. Jiangni et al., Chinese Jour. Of Semics., 27, 1537, (2006).
13. M. Q. Cai et al., Chem. Phys. Lett., 388, 223, (2004).
14. N. Balachandran et al., Jour. Solid State Chem., 39, 351, (1981).
15. J. G. Bednorz et al., Chem. Phys. Letts, 52, 2289, (1984).
16. K. H. Kim et al., J. Phys. Chem. Solids, 46, 1061, (1985).
17. N. Bickel et al., Phys. Rev. Lett., 62, 2009, (1989).
18. W. M. Friedrichs et al., Surf. Sci., 515, 419, (2002).
19. G. Charlton et al., Surf. Sci., 457, L376, (2000).
20. A. Ikeda et al., Surf. Sci., 433, 520, (1999).
21. R. Reihl et al., Phys. Rev. B, 30, 803, (1984).
22. P. Pertosa et al., Phys. Rev. B., 17, 2011, (1978).
23. N. B. Brookes et al., Solid State Commun., 57, 473, (1986).
24. J. Padilla et al., Phys. Rev. B, 56, 1625, (1997).
25. J. Padilla et al., Surf. Sci., 418, 64, (1998).
26. C. Cheng et al., Phys. Rev. B, 62, 10409, (2000).
27. S. Tinte et al., AIP Conf. Proc., 535, 273, (2000).
28. V. I. Anisimov et al., Phys. Rev. B, 44, 943, (1991).
29. V. I. Anisimov et al., Phys. Rev. B, 43, 7570, (1991).
30. V. I. Anisimov et al., Phys. Condens. Mater., 9, 767, (1997).
31. J. K. H. Madsen et al., Europhys. Lett., 69, 777, (2005).
32. J. Heyd et al., J. Chem. Phys., 118, 8207, (2003).
33. J. Heyd et al., J. Chem. Phys., 124, 219906, (2006).

34. J. Heyd et al., J. Chem. Phys., 121, 1187, (2004).
35. F. Ruf et al., APL Mater., 7, 031113, (2019).
36. T. A. Whittle et al., Dalton Trans., 46, 7253, (2017).
37. H. Nasstrom et al., J. Mater. Chem. A, 8, 22626, (2020).